\documentclass[aps,amsmath,amssymb,twocolumn]{revtex4}
\usepackage{graphicx}

\newcommand{\p}{\partial}
\newcommand{\eps}{\epsilon}
\newcommand{\ve}{\varepsilon}
\newcommand{\cE}{{\mathcal E}}

\newcommand{\tr}{\textrm{tr}}

\begin{document}

\title{Universal energy spectrum of tight knots and links in
physics}\thanks{This work is a contribution to ``Numerical methods,
simulations, and computations in knot theory and its applications'',
J.~Calvo, K.~Millet, and E.~Rawdon, eds., World Scientific, Singapore,
2004.}

\author{Roman V. Buniy}
\email{roman.buniy@vanderbilt.edu}
\affiliation{Department of Physics and Astronomy, Vanderbilt
University, Nashville, TN 37235, USA}

\author{Thomas W. Kephart} 
\email{thomas.w.kephart@vanderbilt.edu}
\affiliation{Department of Physics and Astronomy, Vanderbilt
University, Nashville, TN 37235, USA}

\begin{abstract}
We argue that a systems of tightly knotted, linked, or braided flux
tubes will have a universal mass-energy spectrum, since the length of
fixed radius flux tubes depend only on the topology of the
configuration. We motivate the discussion with plasma physics
examples, then concentrate on the model of glueballs as knotted QCD
flux tubes. Other applications will also be discussed.
\end{abstract}

\maketitle

\section {Introduction}

It is known from plasma physics that linked magnetic flux tubes are
much more stable than an unknotted single
loop~\cite{Moffatt:1985}. Linked flux tubes carry topological charge,
and this can be thought of as a conserved (at least to lowest order)
physical quantum number. Similarly, knotted flux tubes carry
topological quantum numbers, and one can think of a knot as a
self-linked loop. The topological charges are described by knot
polynomials that are related to projections of knots or links into a
plane where the crossings of the loops are assigned various
attributes. Following each line around its loop generates the
polynomials. Several types of polynomials have been studied in the
literature (see e.g. Refs.~\cite{Rolfsen,Kauffman}): Alexander,
Conway, Jones, Kauffman, etc., with increasing levels of precision for
distinguishing knots. For example, the simplest knot, the trefoil, has
a chiral partner (mirror image) that is not detected by the simpler
polynomials, but is by the more sophisticated ones. Hence, a pair of
knots with different polynomials are different, but the converse is
not necessarily true. It is still an unsolved problem to find a set of
polynomials that distinguishes all non-isomorphic knots/links. Similar
results hold for braids, and we will also discuss these objects below.

\section {Review of previous physical results on tight knots and links}

If the loops have fixed uniform thickness and circular cross-section
(we will eventually discuss how one can relax this condition), then
each knot and link has a completely specified length if the
configuration is tight, i.e., is of the shortest length with the tubes
undistorted and non-overlapping. If tubes have uniform cross sections,
as can be approximately the case with magnetic or electric flux tubes
carrying quantized flux, or for a polymer or even a piece of
spaghetti, then the length of the tight knot is proportional to the
mass (or energy) of the knot. This, we claim, generates a universal
mass (energy) spectrum for knotted/linked configurations of objects of
this type. The lengths of tight knots were not studied until the
mid-1990s~\cite{bio}, and only recently have accurate calculations of
large numbers of tight knots~\cite{Rawdon} and links~\cite{BKPR}
become available. These results now make it possible to examine
physical systems and compare them with the knot spectrum. The first
physical example studied was tightly knotted DNA~\cite{bio}. More
recently, we have examined the glueball spectrum of
QCD~\cite{Buniy:2002yx}. These particles~\cite{PDG} are likely to be
solitonic states~\cite{Nielsen-Olesen} that are solutions to the QCD
field equations. While QCD will be our main focus in this chapter,
there are many more cases where tight knots may play a role. We first
proceed with an analysis of flux tubes in plasma physics. The lack of
controllable quantum flux renders this case somewhat less interesting
than its generalization to QCD. We will not go into any experimental
details here, but we hope the experts in the areas discussed will take
our general perspective into account when analyzing their data.

In order to decide if a system of flux tubes falls into the universal
class of having a tight knot energy spectrum, we must first
investigate the time scales involved. These are the lifetime of the
soliton $\tau_s$ and the relaxation time $\tau_r$ necessary to reach
the ground state of a tight knot configuration. The soliton lifetime
(or the corresponding decay width $\Gamma_s=1/\tau_s$) can depend on
several factors. These include the effects of flux tube breaking,
rearrangement, and reconnection. The partial width for flux tube
breaking is non-zero if the production of particle/anti-particle pairs
is energetically possible, for example monopole/anti-monopole
($M\bar{M}$) pairs or color monopole/anti-monopole ($M_C \bar{M_C}$)
pairs for magnetic flux (or color magnetic flux) or quark/anti-quark
($q\bar{q}$) pairs for color electric flux tubes. The partial widths
can vary widely depending on the particle masses (e.g., $m_q \ll M$,
so we expect $q\bar{q}$ pairs to be easier to produce than $M\bar{M}$
pairs), interaction strengths (this, for instance, enhances $M\bar{M}$
pair production versus $q\bar{q}$ pair production), and boundary
conditions (tube shape and length). Rearrangement is a quantum effect
where, for example, in a double donut arrangement, the loops can
tunnel free of each other. Finally, reconnection is another effect
where tubes break and re-attach in a different configuration. Such
behavior has been seen in plasma physics, and is of major importance
in understanding a variety of astrophysical systems.  All these
processes change topological charge, and their partial widths compete
more or less favorably with each other depending on the parameters
that describe the system.

\section{Exact calculations}

While no knot lengths have been calculated exactly, it is possible to
calculate the exact lengths of an infinite number of links and many
braids~\cite{BKPR}. For links these calculations are possible in the
case where each individual elements of the link lies in a plane. For
braids, exact calculations are possible when the elements of the braid
are either straight sections or where their centerlines follow helical
paths.  The shortest of all links, the double donut, is exactly
calculable. The two elements lie in perpendicular planes and are tori
of equal length.  The shortest non--trivial braid is a helically
twisted pair.  ``Weyl's tube formula'' is the ideal tool for
calculating the volume of a flux tube~\cite{Weyl,Gray}. The formula
states that for a tube of constant cross-section $\sigma $ normal to a
path of length $l$, the volume of the tube is just $V_T=l\sigma$. This
is a remarkable result that holds in flat 3D. In higher dimensions or
curved space, the result is more complicated, but here we need only
the simple 3D case. This means that if we have an analytic form for
the path and a circular cross-section we can find $V_T$. This leads to
the fact that there is a class of exactly calculable links and braids.
Since there are no known analytic forms for the path of tight knots,
their volumes can only be calculated numerically, but once we have an
estimate of the length, we can then also estimate the volume and
therefore the energy for a corresponding physical system. Braids are
also of physical interest. For the simplest example, a tightly twisted
pair, the path of the center lines of such tubes are helices, and so
the lengths $l(\theta_P,h)$ depend on the pitch angle $\theta_P$, the
radius of the tubes $r$, and the height of the braid $h$. Hence the
volume is $V_T=2\sigma l(\theta_P,h)$ where $\sigma=\pi r^2$. The
volume of triple, etc., helical tight braids can also be found
exactly; however, as with knots, the volumes of topologically
non-trivial tight braids (those where the elements are woven together)
can only be found approximately. (The helical twisted pair can also
become non-trivially knotted/linked by identifying top and bottom
boundaries, but we will not pursue this possibility here.) While the
simple helically twisted braid has a volume that depends on the pitch
angle which can potentially be adjusted by experimental conditions
(see below), the tight knots and links derived from braids have no
such adjustable parameter.

\section{Plasma physics}

Before going on to our main example of QCD, let us stop here to
discuss tight links of flux in electromagnetic plasma. This example is
conceptually somewhat simpler and provides motivation for what is to
come.

Movement of fluids often exhibits topological properties (for a
mathematical review see e.g. Ref.~\cite{Arnold}). For conductive
fluids, interrelation between hydro- and magnetic dynamics may cause
magnetic fields, in their turn, to exhibit topological properties as
well. For example, for a perfectly conducting fluid, the (abelian)
magnetic helicity $\int d ^3 x\,\eps^{ijk}A_i\p_jA_k$ is an invariant
of the motion~\cite{Woltier}, and this quantity can be interpreted in
terms of knottedness of magnetic flux lines~\cite{Moffatt:1969}. Let
us discuss this in some detail since its implications are central to
our more general results.

\subsection{Magnetic relaxation}

A perfectly conducting, viscous and incompressible fluid relaxes to a
state of magnetic equilibrium without a change in
topology~\cite{Moffatt:1985}. The system approaches a state of
magnetic equilibrium by decreasing its magnetic energy, which is
achieved by contraction of the magnetic field lines. In the case of
trivial topology, where field lines are unknotted and unlinked closed
curves that can be contracted to a point without crossing each other,
such magnetic relaxation proceeds uninterrupted. For example, the
initial toroidal field configuration upon contraction deforms into a
configuration of poloidal fields confined to a tube perpendicular to
the original torus. Such configuration is still unstable since small
disturbances augmented by the magnetic pressure lead to the increase
of length of the tube and decrease of its cross-section. The
relaxation eventually leads to a state with zero fields (vacuum).

If, however, the topology of the initial magnetic fields is
non-trivial (for example, when flux tubes are knotted or linked), the
relaxation stops when flux tubes are tightly knotted or linked. This
happens because the ``freeze-in'' condition (see below) forces
topological restrictions on possible changes in field configurations
and so any initial knots and links of field lines remain topologically
unchanged during relaxation. The energy of a final (equilibrium) state
is determined by topology and is bounded from below. One such bound is
proportional to $| {\cal H} |$, where $ {\cal H}$ is the magnetic
helicity (see Ref.~\cite{Moffatt:1985} for details).

\subsection{Abelian helicity}

Consider an abelian gauge potential 1-form $A=A_i d x^i$ and the
corresponding field-strength 2-form $F=\frac{1}{2}F_{ij} d x^i d x^j$.
The helicity for the field inside volume $V$ is defined by
\begin{equation}
{\cal H}=\int_V AF. 
\end{equation}
Under a gauge transformation, $A\to A+ d\psi$ and $F\to F$, and so
using the Bianchi identity $ d F=0$ and the Stokes theorem we find
\begin{equation}
{\cal H}\to {\cal H}+\int_{\p V} \psi F.
\end{equation}
The helicity is thus gauge invariant if the normal component of the
field $F$ vanishes on the surface $\p V$.

It is easy to calculate the helicity for two linked flux tubes with
fluxes $\Phi_1$ and $\Phi_2$. Considering first infinitely thin tubes
(centered around the curves $C_1$ and $C_2$) and integrating over
their cross sections we find 
\begin{equation}
{\cal H}=\Phi_1\int_{C_1}A+\Phi_2\int_{C_2}A.
\end{equation}
The Stokes theorem now leads to 
\begin{equation}
{\cal H}=2n\Phi_1\Phi_2, 
\end{equation}
where $n$ is the Gauss linking number of the two tubes (the algebraic
number of times that one tube crosses the surface spanned by the other
tube). It is straigtforward to generalize this to the case of linked
and/or self-linked thick flux tubes.

\subsection{Non-abelian helicity}

Now we begin to step toward QCD by considering a non-abelian
plasma. Since both $A$ and $F$ are conserved during the plasma motion,
any combination of $A$ and $F$ is a candidate for a conserved quantum
number $ {\cal H}$. The choice of expression is narrowed by requiring
that $ {\cal H}$ is a topological quantity. In particular, this
requires that $ {\cal H}$ be a surface integral. By analogy with the
abelian case, for a conserved non-abelian
helicity~\cite{Jackiw:2000cd}, we choose the corresponding expression
with topological properties,
\begin{equation}
{\cal H}=\int_V \left(A d A+\tfrac{2}{3}A^3\right).
\end{equation}
A (time-independent) gauge transformation $A\to g^{-1}Ag+g^{-1} d g$
leads to
\begin{equation}
{\cal H}\to {\cal H}-\tfrac{1}{3}\tr\int_V \left(g^{-1} d g\right)^3
+\tr\int_{\p V}g d g^{-1}A .\label{cHnA}
\end{equation}
We consider only a limited set of gauge transformations such that do
not change the condition $v^\mu A_\mu=0$; this implies the constraint
$v^i\p_i g=0$. Because of this constraint only two of the components
$g^{-1}\p_i g$ are independent and so $(g^{-1} d g)^3=0$. The helicity
is invariant if the normal component of $A$ vanishes on the surface
$\p V$. This condition is more restrictive than the one needed in the
abelian case and implies the latter.

\subsection{``Freeze-in'' condition}

In a perfectly conductive relativistic non-abelian plasma, the
electric field vanishes in the local frame moving with the plasma. The
Lorentz transformation to the rest frame gives $v^\nu F_{\mu\nu}=0$,
where $v^\mu=(1,v^i)$ is the local plasma velocity and
\begin{equation}
F_{\mu\nu}=\p_\mu A_\nu-\p_\nu A_\mu+[A_\mu,A_\nu]
\end{equation}
is the field strength in terms of the gauge potential $A_\mu$. (The
equation $v^\nu F_{\mu\nu}=0$ is the appropriate generalization of its
familiar abelian non-relativistic counterpart~\cite{electrodynamics}
$E_i=\eps_{ijk}v^jB^k$.) In the ``hydrodynamic gauge'' $v^\mu
A_{\mu}=0$, from the definition of the field strength $F_{\mu\nu}$ we
obtain 
\begin{equation}
\p_0 A_i+(\p_k A_i)v^k+A_k(\p_iv^k)=0\label{A}
\end{equation}
and 
\begin{equation}
\p_0 F_{ij}+(\p_k F_{ij})v^k+F_{ik}(\p_j v^k)+F_{kj}(\p_i
v^k)=0.\label{dF}
\end{equation}

As a fluid particle moves in space, the rate of change of a local
quantity is given by the Lagrange derivative $d/d
t=v^\mu\p_\mu$. Applying this operator to differential forms $A=A_i d
x^i$ and $F=\frac{1}{2}F_{ij} d x^i d x^j$, and using Eqs.~(\ref{A}),
(\ref{dF}) and
\begin{equation}
(d/dt)dx^i=(\p_j v^i)dx^j,
\end{equation}
we find that $A$ and $F$ are constants of motion: 
\begin{eqnarray}
&&dA/dt=0,\\ &&dF/dt=0.
\end{eqnarray}
In other words, for a curve $C$ and a surface $S$ moving with the
fluid, integrals $\int_C A$ and $\int_S F$ do not change in time: the
magnetic lines are ``frozen'' into the plasma. Note the integral
$\int_C A$ is usually called a Wilson loop in the high-energy physics
literature and $\int_S F$ is related to the Wilson loop via Stokes'
theorem.

The freeze-in condition derived in this subsection generalizes the
well-known (see e.g. Ref.~\cite{electrodynamics}) ``freeze-in''
condition in MHD to non-abelian case. The non-relativistic analog of
this result was obtained in Ref.~\cite{Holm:1984yh}.

\section {QCD and Glueballs}

\subsection{QCD}

What is the ideal physical system in which to discover and study tight
knots and links? We claim it is Quantum Chromodynamics (QCD), so in
order to explain our reasoning we first pause to briefly summarize
QCD~\cite{PDG}.

To lowest order, the standard model of particle physics can be broken
into two major sectors, one describing the electroweak processes for
leptons, quarks, photons and intermediate vector bosons, like
electron-neutrino scattering, and the other describing the strong
interactions. The electroweak processes are easier to deal with
because the coupling constants of this sector are small (e.g., the
fine structure constant $\alpha = e^2/{4\pi}$ is approximately
$1/137$) so that perturbative calculations can be preformed to a high
precision and are in beautiful agreement with experiment. On the other
hand, the strong interactions which describes the interactions of
quarks and gluons, while in principle completely described by QCD, are
much more difficult to deal with because at low energy the coupling
constant $\alpha_S = {g_S}^2/{4\pi}$ is $O(1)$. This, for instance,
makes the quark-antiquark ($q\bar{q}$) bound state problem
analytically intractable.

In somewhat more detail, the QCD is a gauge theory with the lagrangian
density
\begin{equation}	
{\cal L}=-\tfrac {1}{4} F^a_{\mu \nu }F^{a \mu
\nu}+\bar{\psi}^i_q\gamma^\mu (D_\mu )_{ij}{\psi}^j_q
-m_q\bar{\psi}^i_q{\psi}^i_q,
\label{eq:1}
\end{equation}
where the field strength is
\begin{equation}	
  F^a_{\mu \nu } =\partial_\mu A^a_\nu + \partial_\nu A^a_\mu
  -g_Sf_{abc}A^b_\mu A^c_\nu
\label{eq:2}
\end{equation}
and the covariant derivative is
\begin{equation}	
(D_\mu )_{ij}=\delta_{ij}\partial_\mu +
\tfrac{1}{2}\lambda^a_{ij}A^a_\mu.
\label{eq:3}
\end{equation}

Here the ${\psi}^j_q$s are the Dirac spinor quark fields with color
($i$) and flavor ($q$) indices, the Yang-Mills fields $A^a_\mu$
(connection) describe the gluons, the $f_{abc}$ are the structure
constants for the $SU(3)$ gauge group, and the $\lambda^a_{ij}$'s are
the triplet representation matrix elements for the quarks. Quarks come
in six flavors with six associated quantum numbers conserved by the
strong interactions but all except electric charge can be violated by
the electroweak interactions. Color charges are confined due to the
fact that $SU(3)$ is an unbroken symmetry with asymptotic
freedom. I.e., the strong coupling $\alpha_S$ is energy dependent and
becomes small at high energy but is large at low energies where it
becomes $O(1)$ at a few hundred MeV. Then the theory becomes
non-perturbative and confining (color charges cannot be isolated) with
all observables color singlet states. These singlets are described by
the quark model and are either bosonic $q\bar{q}$ bound state called
mesons (pions, kaons, etc.)  or fermionic $qqq$ states called baryons
(protons, neutrons, etc.). Besides the mesons and baryons there are a
number states that do not fit neatly into the quark model. These go by
the names of hybrid states, exotic states and glueballs. Hybrid states
and exotic states are thought to be unusual combinations of quarks
e.g., $qq\bar{q}\bar{q}$ bosonic states, or $qqqq\bar{q}$ fermionic
states, most of which have net flavor charges. The glueballs on the
other hand are thought to be made of gluons with at most some virtual
quark content, hence no flavor charges.

We are now in a position to support our claim that QCD is the ideal
physical system in which to discover and study tight knots and
links. Here are our reasons: 
\begin{enumerate}
\item QCD is a solidly based part of the standard model of particle
physics, and much about color confinement and the quark model is
already well understood in this context, making much previous work
transferable to the problem of tightly knotted flux tubes in this
theory.
\item Unlike plasmas, fluids or other condensed matter systems where
flux tubes are excitations of some media with many parameters that
could hide universal behavior, flux knots in QCD can exist in the
vacuum. Thus continuum states are absent and there are no media
parameters to vary and obscure the universality. Hence, the results in
QCD can be far less ambiguous.
\item The hadronic energy spectrum has been measured over a large
range of energies (140 MeV to 10 GeV) and already many hundreds of
states are known. We expect that among these, a few dozen can be
classified as tightly knotted/linked flux tubes states. These states
must have no valance quarks (i.e., no flavor quantum numbers) in order
to be classified as glueballs.
\item Knotted solitons in QFT are already known to exist. 
\item One can efficiently search for new glueball states at
accelerators. (Also, data from older experiments still exist and can
be reanalyzed to check the predictions of new states described below.)
\end{enumerate}

\subsection{Knot energies}

Consider a hadronic collision that produces some number of baryons and
mesons plus a gluonic state in the form of a closed QCD flux tube (or
a set of tubes). From an initial state, the fields in the flux tubes
quickly relax to an equilibrium configuration, which is topologically
equivalent to the initial state. (We assume topological quantum
numbers are conserved during this rapid process.) The relaxation
proceeds through minimization of the field energy. Flux conservation
and energy minimization force the fields to be homogeneous across the
tube cross sections. This process occurs via shrinking the tube
length, and halts to form a ``tight'' knot or link. The radial scale
will be set by $\Lambda _{\textrm{QCD}}^{-1}$. The energy of the final
state depends only on the topology of the initial state and can be
estimated as follows. An arbitrarily knotted tube of radius $a$ and
length $l$ has the volume $\pi a^2 l$. Using conservation of flux
$\Phi_E$, the energy becomes $\propto l(\tr\Phi_E^2)/(\pi
a^2)$. Fixing the radius of the tube (to be proportional to
$\Lambda_{\textrm{QCD}}^{-1}$), we find that the energy is
proportional to the length $l$. The dimensionless ratio
$\ve(K)=l/(2a)$ is a topological invariant and the simplest definition
of the ``knot energy''~\cite{knot_energy}.

\begin{table*}
\caption{\label{table}Comparison between the glueball mass spectrum
and knot energies.}
\begin{ruledtabular}
\begin{tabular}{ccccc} State & Mass & 
$K$~\footnote{Notation $n^l_k$ means a link of $l$ components with $n$
crossings, and occurring in the standard table of links (see
e.g. \protect\cite{Rolfsen}) on the $k^\textrm{th}$ place. $K\#K'$
stands for the knot product (connected sum) of knots $K$ and $K'$ and
$K*K'$ is the link of the knots $K$ and $K'$.} &
$\ve(K)$~\footnote{Values are from \protect\cite{bio} except for our
exact calculations of $2^2_1$, $2^2_1*0_1$, and $(2^2_1*0_1)*0_1$ in
square brackets, our analytic estimates given in parentheses, and our
rough estimates given in double parentheses.} &
$E(G)$~\footnote{$E(G)$ is obtained from $\ve(K)$ using the fit
(\ref{fit1}).} \\ \hline {\rule[1mm]{0mm}{3mm} $f_0(600)$} &
$400-1200$ & $2^2_1$ & $12.6\ [4\pi]$ & $768\ [766]$\\ $f_0(980)$ &
$980\pm 10$ & $3_1$ & $16.4$ & $993$\\ $f_2(1270)$ & $1275.4\pm 1.2$ &
$2^2_1*0_1$ & $[6\pi+2]$ & $[1256]$\\ $f_1(1285)$ & $1281.9\pm 0.6$ &
$4_1$ & $21.2$ & $1277$\\ & & $4^2_1$ & $(21.4)$ & $(1289)$\\
$f_1(1420)$ & $1426.3\pm 1.1$ & $5_1$ & $24.2$ & $1454$\\$\{f_2(1430)$
& $\approx 1430\}$~\footnote{States in braces are not in the Particle
Data Group (PDG) summary tables.} & $5_1$ & $24.2$ & $1454+\delta'$\\
$f_0(1370)$ & $1200-1500$ & $3_1*0_1$ & $(24.7)$ &
$(1484)$\\$f_0(1500)$ & $1507\pm 5$ & $5_2$ & $24.9$ & $1496$\\
$\{f_1(1510)$ & $1518\pm 5\}$ & $5_2$ & $24.9$ &
$1496+\delta$\\$f'_2(1525)$ & $1525\pm 5$ & $5_2$ & $24.9$ &
$1496+3\delta$\\ $\{f_2(1565)$ & $1546\pm 12\}$ & $5^2_1$ & $(25.9)$ &
$(1555)$\\ $\{f_2(1640)$ & $1638\pm 6\}$ & $6^3_3$ & $((27.3))$ &
$((1638))$\\ \multicolumn{5}{c}{.\dotfill.}\\ & &
$(2^2_1*0_1)*0_1$~\footnote{This is the link product that is not
$2^2_1*2^2_1$.} & $[8\pi+3]$ & $[1686]$~\footnote{Resonances have been
seen in this region, but are unconfirmed~\cite{PDG}.}\\$f_0(1710)$ &
$1713\pm 6$ & $6^3_2$ & $((28.6))$ & $((1714))$\\ & & $3_1\#3_1^*$ &
$28.9\ (30.5)$ & $1732\ (1827)$\\ & & $3_1\#3_1$ & $29.1\ (30.5)$ &
$1744\ (1827)$\\ & & $2^2_1*2^2_1$ & $[8\pi+4]$ & $[1745]$\\ & & $6_2$
& $29.2$ & $1750$\\ & & $6_1$ & $29.3$ & $1756$\\ & & $6_3$ & $30.5$ &
$1827$\\ & & $7_1$ & $30.9$ & $1850$\\ & & $8_{19}$ & $31.0$ &
$1856$\\ & & $8_{20}$ & $32.7$ & $1957$\\ $f_2(2010)$ &
$2011^{+60}_{-80}$ & $7_2$ & $33.2$ & $1986$\\ $f_4(2050)$ & $2025\pm
8$ & $8_{21}$ & $33.9$ & $2028$\\ & & $8_1$ & $37.0$ & $2211$\\ & &
$10_{161,162}$ & $37.6$ & $2247$\\ $f_2(2300)$ & $2297\pm 28$ &
$8_{18}$, $9_1$ & $38.3$ & $2288$\\ $f_2(2340)$ & $2339\pm 60$ & $9_2$
& $40.0$ & $2389$\\ & & $10_1$ & $44.8$ & $2672$\\ & & $11_1$ & $47.0$
& $2802$\\
\end{tabular}
\end{ruledtabular}
\end{table*}

Many knot energies have been calculated by Monte Carlo
methods~\cite{bio} and certain types can be calculated exactly (see
below), while for other cases simple estimates can be made (see
Table~\ref{table}). For example, the knot energy of the connected
product of two knots $K_1$ and $K_2$ satisfies
\begin{equation}
\varepsilon(K_1\#K_2)<\varepsilon(K_1)+\varepsilon(K_2).
\end{equation} 
A rule of thumb is
\begin{equation}
\varepsilon(K_1\#K_2)\approx
\varepsilon(K_1)+\varepsilon(K_2)-(2\pi-4).
\end{equation}

\begin{figure*}
\centering \includegraphics[angle=0,width=9cm]{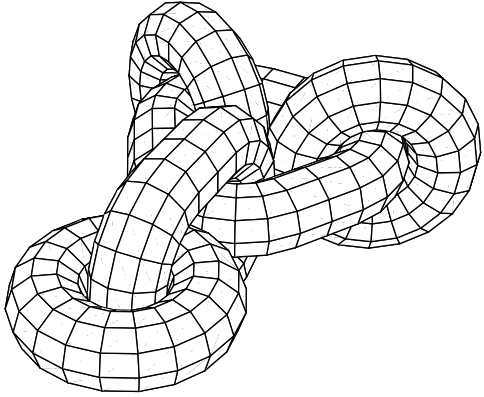}
\caption{\label{exact} An example of an exactly calculable link.}
\end{figure*}

Most of the knot energies in Table 1 have been taken from
Ref.~\cite{bio}, but we have independently calculated the energy of
$2^2_{1}$, $4^3_1$ and $6^4_1$ exactly and the energy for several
other knots and links approximately. We find
$\varepsilon(2^2_{1})=4\pi\approx 12.57$, to be compared with the
Monte Carlo value $12.6$. We also find $\varepsilon(4^3_{1})=6\pi+2$
and $\varepsilon(6^4_{1})=8\pi+3$, where there are no Monte Carlo
comparisons available, or needed. Other exactly calculable links can
be found in Ref.~\cite{BKPR} and an example of a link with energy
$10\pi+5$ is shown in Fig.~\ref{exact}.

\subsection{Model}

In our model, the chromoelectric fields $F_{0i}$ are confined to
knotted/linked tubes. After an initial time evolution, the system
reaches a static equilibrium state which is described by the energy
density
\begin{equation}
{\cE}_E=\tfrac{1}{2}\tr F_{0i}F^{0i}-V.\label{L}
\end{equation}
Similar to the MIT bag model~\cite{MIT-bag}, we have included a
constant potential energy $V$ needed to keep the tubes at a fixed
cross-section. The chromoelectric flux $\Phi_E$ is conserved and we
assume flux tubes carry one flux quantum. To account for conservation
of the flux, we add to Eq.~(\ref{L}) the term
\begin{equation}
\tr\lambda\{\Phi_E/(\pi a^2)-n^iF_{0i}\},\label{flux1}
\end{equation}
where $n^i$ is the normal vector to a section of the tube and
$\lambda$ is a Lagrange multiplier. The energy density should be
constant under variations of the degrees of freedom, the gauge
potentials $A_\mu$. This leads to the system of equations
\begin{eqnarray}
&&D^0(F_{0i}-\lambda n_i)=0,\\ &&D^i(F_{0i}-\lambda
n_i)=0,\label{flux2}
\end{eqnarray}
which has the constant field
\begin{equation}
F_{0i}=(\Phi_E/\pi a^2)n_i\label{F}
\end{equation}
as its solution. With this solution, the energy is positive and
proportional to $l$ and thus the minimum of the energy is achieved by
shortening $l$, i.e., tightening the knot.

The case of chromomagnetic flux tubes can be similarly
considered. This requires the confinement of color magnetic flux tubes
which is possible if there are light quarks in the spectrum of the
theory~\cite{Goldhaber:1999sj}.

Lattice calculations, QCD sum rules, electric flux tube models, and
constituent glue models agree that the lightest non--$q\bar{q}$ states
are glueballs with quantum numbers $J^{++}=0^{++}$ and $2^{++}$ (see
Ref.~\cite{West}). We will model all $J^{++}$ states (i.e., all
$f_{J}$ and $f'_J$ states listed by the PDG~\cite{PDG}), some of which
will be identified with rotational excitations, as knotted/linked
chromoelectric QCD flux tubes. We proceed to identify knotted and
linked QCD flux tubes with glueballs, where we include all $f_J$ and
$f'_J$ states. The lightest candidate is the $f_0(600)$, which we
identify with the shortest knot/link, i.e., the $2^2_1$ link; the
$f_0(980)$ is identified with the next shortest knot, the $3_{1}$
trefoil knot, and so forth. All knot and link energies have been
calculated for states with energies less then
$1680\,\textrm{MeV}$. Above $1680\,\textrm{MeV}$ the number of knots
and links grows rapidly, and few of their energies have been
calculated. However, we do find knot energies corresponding to all
known $f_J$ and $f'_J$ states, and so can make preliminary
identifications in this region. (We focus on $f_J$ and $f'_J$ states
from the PDG summary tables. The experimental errors are also quoted
from the PDG. There are a number of additional states reported in the
extended tables, but some of this data is either conflicting or
inconclusive.)

Our detailed results are collected in Table \ref{table}, where we list
$f_J$ and $f'_J$ masses, our identifications of these states with
knots and the corresponding knot energies.

\begin{figure*}
\centering \includegraphics[angle=0,width=11.4cm]{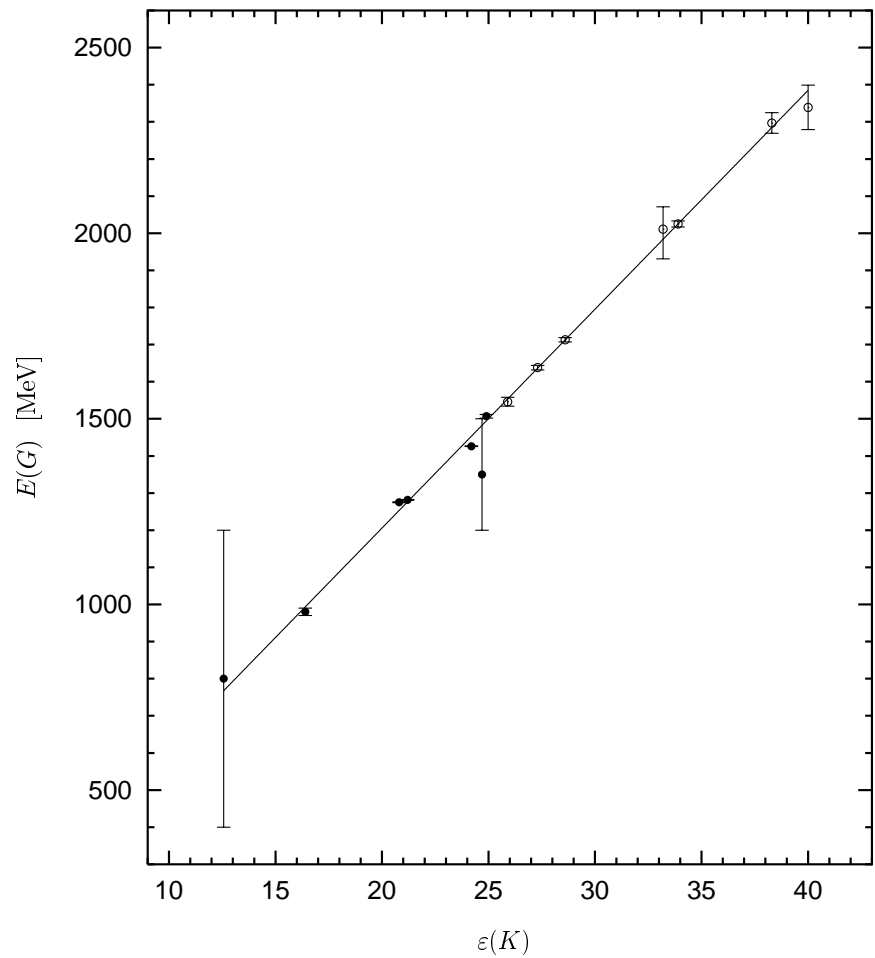}
\caption{\label{figure0}Relationship between the glueball spectrum
$E(G)$ and knot energies $\varepsilon(K)$. Each point in this figure
represents a glueball identified with a knot or link. The straight
line is our model and is drawn for the fit (\ref{fit1}).}
\end{figure*}

\begin{figure*}
\centering \includegraphics[angle=0,width=11.4cm]{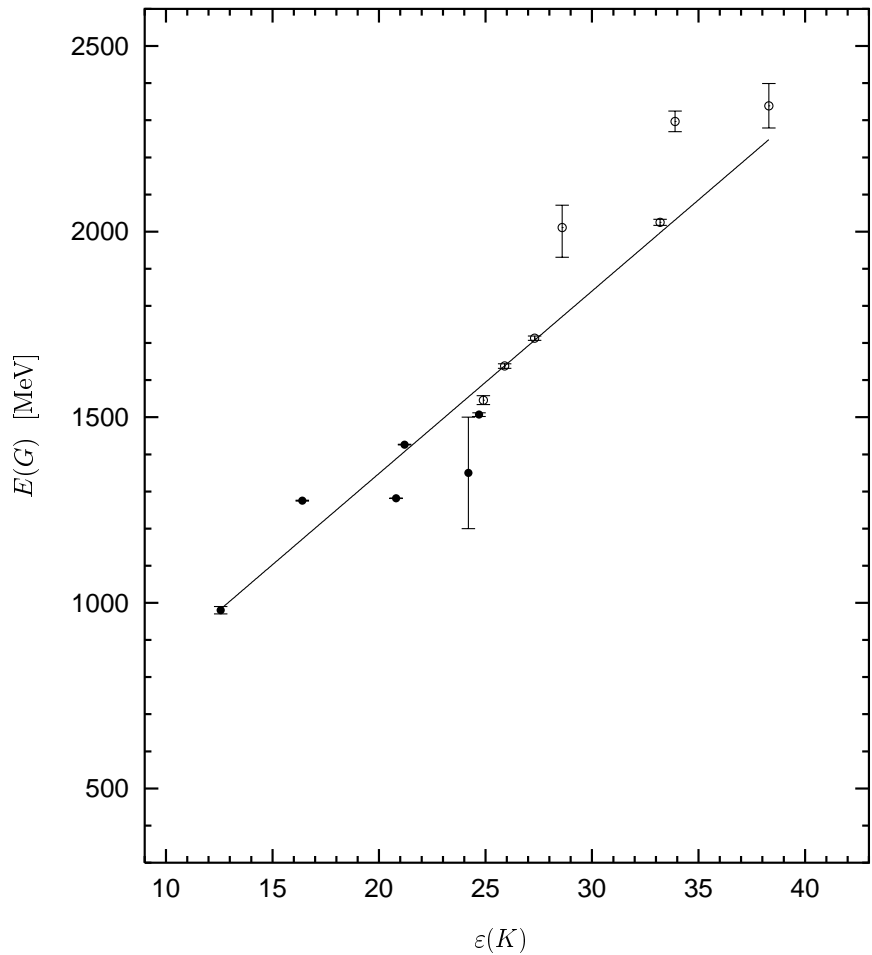}
\caption{\label{figure1}The same as in Fig.~\ref{figure0}, but the
first glueball is missed out.}
\end{figure*}

\begin{figure*}
\centering \includegraphics[angle=0,width=11.4cm]{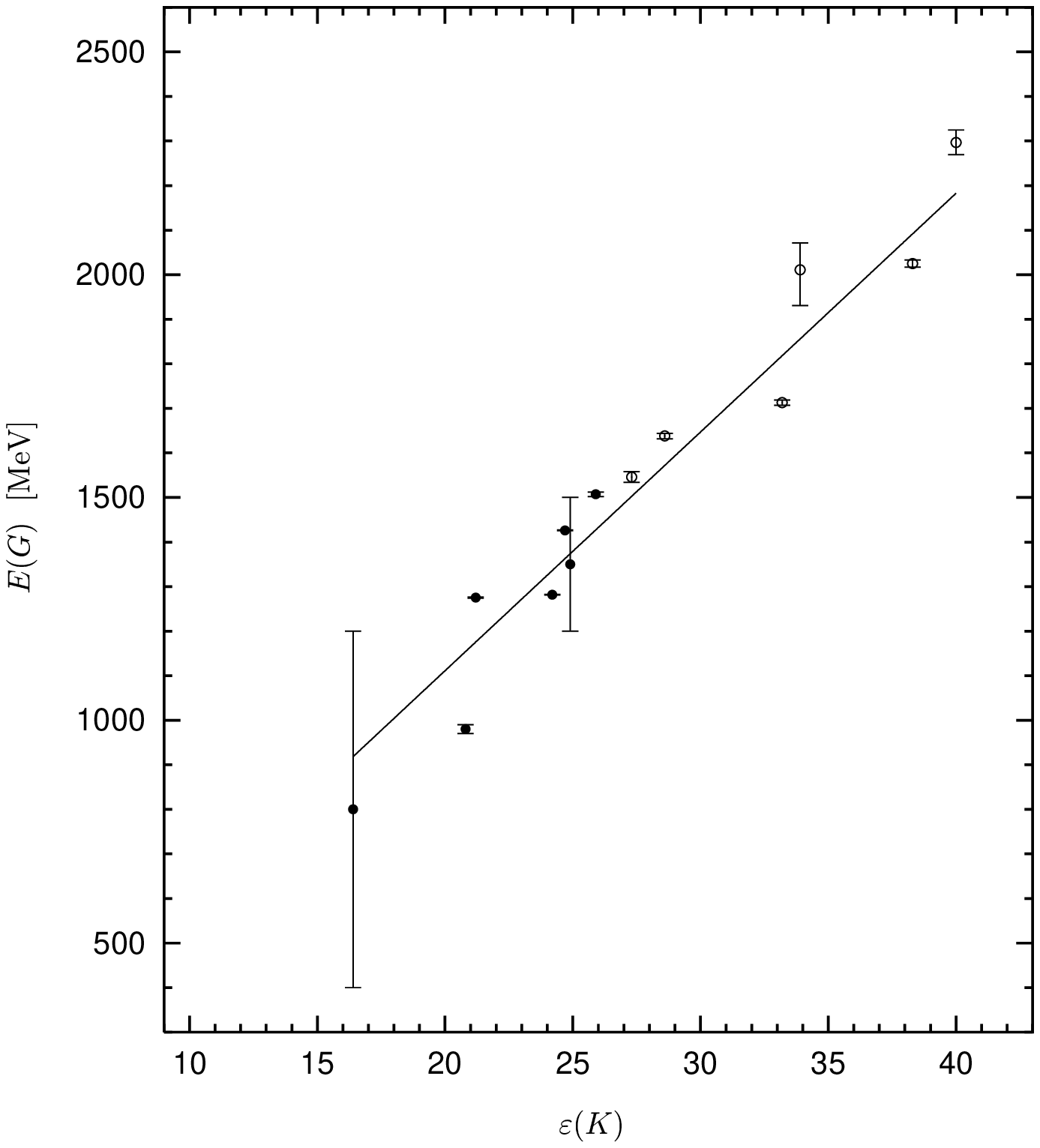}
\caption{\label{figure2}The same as in Fig.~\ref{figure0}, but the
first knot/link is missed out.}
\end{figure*}

In Fig.~\ref{figure0} we compare the mass spectrum of $f_J$ states
with the identified knot and link energies. Since errors for the knot
energies in Ref.~\cite{bio} were not reported, we conservatively
assumed the error to be $1\%$. A least squares fit to the most
reliable data (below $1680\,\textrm{MeV}$) gives
\begin{equation}
E(G)=(23.4\pm
46.1)+(59.1\pm 2.1)\varepsilon(K)\ \ \ [\textrm{MeV}],\label{fit1}
\end{equation} 
with $\chi^2=9.1$. The data used in this fit is the first seven $f_J$
states (filled circles in Fig.~\ref{figure0}) in the PDG summary
tables. Inclusion of the remaining seven (non-excitation) states
(unfilled circles in Fig.~\ref{figure0}) in Table~\ref{table}, where
either the glueball or knot energies are less reliable, does not
significantly alter the fit and leads to
\begin{equation}
E(G)=(26.9\pm 24.9)+(58.9\pm 1.0)\varepsilon(K)\ \ \
[\textrm{MeV}],\label{fit2}
\end{equation} 
with $\chi^2=10.1$. The fit (\ref{fit1}) is in good agreement with our
model, where $E(G)$ is proportional to $\varepsilon(K)$. Better HEP
data and the calculation of more knot energies will provide further
tests of the model and improve the high mass identification.

For comparison, we have fitted the same data in Fig.~\ref{figure1}
except that the first glueball is missed out; this results in $\chi
^{2}=231$. Similarly, we have fitted the same data in
Fig.~\ref{figure2} except that this time the first link is missed out;
this results in $\chi^{2}=355$. This is strong evidence that our
identification is appropriate.

In terms of the bag model~\cite{MIT-bag}, the interior of tight knots
correspond to the interior of the bag. The flux through the knot is
supported by current sheets on the bag boundary (surface of the
tube). Knot complexity can be reduced (or increased) by unknotting
(knotting) operations~\cite{Rolfsen,Kauffman}. In terms of flux tubes,
these moves are equivalent to reconnection
events~\cite{reconnection}. Hence, a metastable glueball may decay via
reconnection. Once all topological charge is lost, metastability is
lost, and the decay proceeds to completion. Two other glueball decay
processes are: flux tube (string) breaking; this favors large decay
widths for configurations with long flux tube components; and quantum
fluctuations that unlink flux tubes; this would tend to broaden states
with short flux tube components. As yet we are not able to go beyond
providing a phenomenological fit to these qualitative
observations~\cite{Buniy:2002yx}, but hope to be able to do so in the
future.

We have assumed one fluxoid per tube. There may be states with more
than one fluxoid, but these would presumably have somewhat fatter flux
tubes with higher flux densities and higher energies. For example, the
two fluxoid trefoil knot $3_1$ would certainly have
$\varepsilon(K)>2\,\varepsilon(3_1)$ and a fairly reliable estimate
gives $\varepsilon(K)\approx 2\sqrt{2}\,\varepsilon(3_1)$. Hence most
multifluxoid states would be above the mass range of known glueball
candidates.

\section{Discussion and conclusions}

We expect all tight knots and links to be described by smooth curves
with minimum radius of curvature $2r$ where $r$ is the tube
radius. Consider how we would approximate such a curve on a square
lattice. The simplest nontrivial example is the double donut. We
consider one of its loops (a circle) which we call $\Gamma$ to be
centered on a particular lattice point $P$. The radius of the circle
$R$ is some number of lattice spacings away from $P$ (we assume we are
in a lattice plane), along the direction of one of $P$'s nearest
neighbors. Assume all other points on $\Gamma$ are at a distance $d
\geq R$ away from $P$. If we minimize the length $L$ of $\Gamma$, then
the ratio of $L$ to the circumference of a circle of radius $R$ is
\begin{equation}
\frac {L_{\textrm{min}}}{2\pi R}=4/\pi,
\end{equation}
which is the best approximation (about a 27\% error) we can have on a
square lattice. Similar arguments show we can do somewhat better on a
hexagonal lattice where we can achieve
\begin{equation}
\frac {L_{\textrm{min}}}{2\pi R}=2\sqrt{3}/\pi,
\end{equation}
about a 10\% error.  However, neither of these approximations are
stunningly successful, nor does decreasing the lattice spacing improve
the approximations. We assume other knots and links will be
approximated on the lattice with similar accuracy. It is interesting
to note here that both lattice estimates are too high and that lattice
QCD typically predicts glueball masses above 1 GeV, while Monte Carlo
predictions for tight knots leads to a lighter glueball spectrum with
lightest state just below 800 MeV. Lattice QCD are certainly more
sophisticated and involved than the simple estimates we have just
made, but we do expect requiring knots/links to lie on a lattice to
give glueball masses on the high side. Furthermore, the tube cross
section on a lattice is not circular, inhibiting tight packing, and
the amount of curvature energy on the lattice is likely to be more
than for a smooth curve. We expect these effect to also contribute to
increasing the lattice glueball mass estimates.

Let us return to continuum physics and consider a slab of material
that can support flux tubes. We have in mind a superfluid or
superconductor, but are not limited to these possibilities. Assume
further that the flux tubes carry one and only one unit of flux. Next
consider manipulating these flux tubes. For instance, consider a
hypothetical superconductor where the flux tubes are pinned at the
bottom of the slab, say by being attracted to the poles of some
magnetic material, and at the top of the slab they are each associated
with the pole of a movable permanent magnet, perhaps a magnetic
whisker, or fine solenoid. If the north poles of a flux tube is on the
top surface, we call it a ($+$) flux tube (and ($-$) for the south
pole at the top). Since the flux tubes are pinned on the bottom of the
slab we can maneuver the top ends into a braid. For example, if there
are two flux tubes we can rotate them around each other to make a
helically twisted pair. If there are three or more flux tubes we can
twist them or braid them. If we measure the forces on the magnets when
the flux tubes are being manipulated and use these to find the energy
needed to form a braided configuration, then we gain information on
the tension of the flux tubes and on the energy stored in a tight
braid.  If we have both ($+$) and ($-$) flux tubes we can bring ($+$)
-- ($-$) pairs together, annihilate the ends so the tube can be pulled
into the interior of the superconductor. If we do this at both the top
and bottom surfaces for a braided pair then it should be possible to
form a bulk knot (similar for links). Assuming they have time to relax
to tight configurations, the energy released by the eventual decay of
these structures in the bulk should correspond to the universal energy
spectrum described above.  Perhaps there is no system with all the
properties of our ideal superconductor; however, there are systems
that contain many if not most of the desired features.

Another collection of physical systems of potential interest for its
ability to support vortices, and so knots and braids, are the atomic
Bose-Einstein condensates. For example, dilute ${ }^{87}$Rb atoms at
$80\ \textrm{nK}$. Recently results have been presented demonstrating
laser stirring of these condensates to produce vortices~\cite{BEC}. As
many as seven vortices have been seen in a $0.1\ \textrm{mm}^2$
region. One could hope that advancement in these techniques could lead
to more complicated solitonic structures, in particular knots, links,
and braids. The advantage of these systems is the experimental
accessibility of these structures, and the large percentage of the
volume of the system taken up by the solitons. This could potentially
dramatically improve the signal-to-noise ratio in the study of tight
knots in condensed matter systems.

We have argued that knotted/linked solitonic physical systems have
universal mass-energy spectra and have demonstrated this with a
detailed example from QCD. We have provided other examples that are
good candidates but where the universality has yet to be seen.  A
system is a candidate if it contains line solitons that can (somewhat
paradoxically) relax to their tight knot/link ground state in a time
shorter than their decay time. Future work along these lines could tie
knot theory to many subfields of physics or to other sciences.

\end{document}